# Conditional regression based on a multivariate zero-inflated logistic normal model for microbiome relative abundance data

Zhigang Li[1,2,3*], Katherine Lee[4], Margaret R. Karagas[2,3], Juliette C. Madan[2,3,5], Anne G. Hoen[1,2,3], A. James O'Malley[1,6], Hongzhe Li[7]

[1]Department of Biomedical Data Science, Geisel School of Medicine at Dartmouth, 1 Medical Center Drive, Lebanon, NH 03756, USA, [2]Children's Environmental Health and Disease Prevention Research Center at Dartmouth, Hanover, New Hampshire, [4]Department of Epidemiology, Geisel School of Medicine at Dartmouth, 1 Medical Center Drive, Lebanon, NH 03756, USA, [4]Phillips Exeter Academy, Exeter, NH 03833, USA, [5]Division of Neonatology, Department of Pediatrics, Children's Hospital at Dartmouth, Lebanon, New Hampshire, [6]The Dartmouth Institute for Health Policy and Clinical Practice, Geisel School of Medicine at Dartmouth, 1 Medical Center Drive, Lebanon, NH 03756, USA and [7]Department of Biostatistics and Epidemiology, University of Pennsylvania School of Medicine, Philadelphia, PA 19104, USA

*To whom correspondence should be addressed. Zhigang.Li@dartmouth.edu

**Abstract** The human microbiome plays critical roles in human health and has been linked to many diseases. While advanced sequencing technologies can characterize the composition of the microbiome in unprecedented detail, it remains challenging to disentangle the complex interplay between human microbiome and disease risk factors due to the complicated nature of microbiome data. Excessive numbers of zero values, high dimensionality, the hierarchical phylogenetic tree and compositional structure are compounded and consequently make existing methods inadequate to appropriately address these issues. We propose a multivariate two-part zero-inflated logistic normal (MZILN) model to analyze the association of disease risk factors with individual microbial taxa and overall microbial community composition. This approach can naturally handle excessive numbers of zeros and the compositional data structure with the discrete part and the logistic-normal part of the model. For parameter estimation, an estimating equations approach is employed that enables us to address the complex inter-taxa correlation structure induced by the hierarchical phylogenetic tree structure and the compositional data structure. This model is able to incorporate standard regularization approaches to deal with high dimensionality. Simulation shows that our model outperforms existing methods. Our approach is also compared to others using the analysis of real microbiome data.

## 1 Introduction

The human microbiome is composed of the collective genomes of commensal, symbiotic and pathogenic microorganisms including bacteria, archaea, viruses, and fungi and is an important contributor to human physiology and disease [1-3]. Perturbation of the microbiome homeostasis or changes in individual microbes have been linked to a variety of human diseases including asthma, infection, and allergy in children [4-6], as well as cancer [7-9] and obesity [10, 11]. High-throughput sequencing technologies such as shotgun metagenomic sequencing and 16s ribosomal RNA gene sequencing have recently been applied to quantify microbes constituting the microbiome [12, 13]. Sequencing reads are usually aligned to known reference sequences [14, 15] in order to identify and quantify the abundance of microbial taxa.

While sequencing technologies can characterize the composition of microbiome in unprecedented detail, it remains challenging to examine the associations of disease risk factors or health outcomes with microbiome data due to the complicated structure of microbiome sequencing data [16]. First, because of enormous between-subject variation in sequencing reads, microbiome data is usually summarized as *relative abundance* (RA) at a certain taxonomy level: essentially the percentage of sequencing reads for each taxon in the sample. Thus, the RA has a compositional structure with the constraint that all the RA must sum to one. Compositional data structure could induce spurious relationships due to the linear dependence between compositional components because an increase in one component must induce a decrease in another component. Second, there is an underlying hierarchical structure of the microbiome data reflecting the evolutionary relationships (phylogeny) between microbes. This hierarchical structure could introduce dependence among taxa on top of the compositional structure. Third, there are excessive numbers of zero sequencing reads for many taxa. This sparsity causes modeling issues for many traditional approaches. Fourth, microbiome data can be of extremely high dimensions because a single sample can produce millions of sequencing reads. Since all of these features arise simultaneously, they are compounded and thus make the analysis of microbiome data much more complicated in practice.

Existing approaches remain inadequate to fully address the modeling challenges when studying the microbiome and its relationships with other variables of interest. Community level metrics of overall diversity such as Simpson index, phylogenetic diversity, and UniFrac distance [17, 18] reduce the dimension of the microbiome data dramatically and thus have straightforward interpretations. This type of methods is not able to decipher the associations of individual microbial taxa with other variables due to dimension reduction prior to association analysis. Differential abundance analysis is useful to compare microbial composition between two groups or multiple groups [19, 20], however, it cannot adjust for covariates which could be important in the presence of confounders. Regression models have also been developed in the literature and can be roughly divided into two categories by how microbiome data is treated in the model: 1) predictors or 2) outcomes. For the first type of models [21-23], RA data are usually used and special handling is needed to deal with the large dimensional and compositional features of the RA data. Zero sequencing reads are often imputed with the pseudo count (ie, 0.5) representing the maximum rounding error. There are two subcategories for the second type of models according to what type of microbiome data is used: a) absolute abundance (ie, sequencing read counts) or b) RA. When modeling absolute abundance data [24-26], overdispersion needs to be appropriately handled and challenges from zero-inflated and high dimensional structures have not been fully addressed in this case. When modeling RA data [27], individual taxa are usually analyzed one by one with a multiple testing correction procedure to control for type I error rate. This approach is not able to incorporate the inter-taxa correlation. Under this setting, there are also methods developed for examining associations between longitudinal microbiome data and clinical covariates [28].

In this paper, we will develop a statistical regression model to identify the associations of disease risk factors with the distribution of microbial taxa. Therefore, microbial RA data form a multivariate dependent variable. A zero-inflated logistic normal model will be proposed to account for the zero-inflated data structure and the compositional structure. We will borrow ideas from GEE [29] to handle the overall correlations between microbiome taxa induced by the compositional structure and hierarchical phylogenic structure. Regularization approaches such as LASSO [30], SCAD [31] and MCP [32] will be incorporated in the method to address high dimensionality of the data. Simulation results show that our approach outperforms existing methods. A real study example is presented to identify infant gut microbial taxa that are associated with environmental exposures in the New Hampshire Birth Cohort Study [33]. All the simulations and real data analyses were done in R.

## 2 A multivariate zero-inflated logistic-normal model and regression

### 2.1 Multivariate logistic-normal distribution

Suppose there are $K + 1$ microbial taxa and let $Y^* = (Y_1^*, \ldots, Y_{K+1}^*)^T$ denote the true relative abundance (RA) of microbial taxa where the sup-script $T$ denotes the transpose of a vector (or matrix). In this section, we don't consider taxa have zero RA for illustration purpose. The RA has a compositional structure with $\sum_{k=1}^{K+1} Y_k^* = 1$ and the vector $Y^*$ lies in the $K$-dimensional simplex $\mathcal{S}^K$ where there are only $K$ degrees of freedom for the $K + 1$ RA's [16].

We first present an brief introduction of the multivariate logistic-normal distribution that has been discussed in the literature [34] and has been proposed for modeling the compositional data. We say that a vector $Y^*$ follows a multivariate logistic-normal (LN) distribution [34, 35] and thus its log-ratio transformation, a $K$-dimensional vector, $U = \left(\log\left(\frac{Y_1^*}{Y_{K+1}^*}\right), \ldots, \log\left(\frac{Y_K^*}{Y_{K+1}^*}\right)\right)^T \triangleq (U_1, \ldots, U_K)^T$ follows a multivariate normal distribution $N(\mu, \Sigma)$ where $\mu = (\mu_1, \ldots, \mu_K)^T$ is the $K$-dimensional mean vector and $\Sigma$ is the $K \times K$ variance matrix.

For any subset of RA's, denoted by $Y_{k_1}^*, \ldots, Y_{k_L}^*, 1 \leq k_1 < \cdots < k_L \leq K + 1, 1 \leq L \leq K + 1$, we can form a subcomposition by recalculating RA's within this subset: $\left(\frac{Y_{k_1}^*}{\sum_{l=1}^L Y_{k_l}^*}, \ldots, \frac{Y_{k_L}^*}{\sum_{l=1}^L Y_{k_l}^*}\right)$. It is straightforward to see that the log-ratio transformation of the subcomposition is a linear transformation of $U$ given by

$$U_{k_1,\ldots,k_L} = \left(\log\left(\frac{Y_{k_1}^*}{Y_{k_L}^*}\right), \ldots, \log\left(\frac{Y_{k_{L-1}}^*}{Y_{k_L}^*}\right)\right)^T = AU,$$

where $A$ is a $(L - 1) \times K$ matrix with the $k_L$th column being -1's, the $(l, k_l)$th elements, $l = 1, \ldots, L - 1$, being 1's and all other elements being zero. If $k_L = K + 1$, then matrix $A$ has the $(l, k_l)$th elements being 1's and all other elements being zero's. So $U_{k_1,\ldots,k_L}$ has a multivariate normal distribution with mean $A\mu$ and variance $A\Sigma A^T$. Therefore, any subcomposition follows a LN distribution as well.

## 2.2 Multivariate zero-inflated logistic-normal distribution

In practice, many taxa may not be observed due to biological conditions. Thus, the observed RA vector $Y = (Y_1, \ldots, Y_{K+1})^T$ is usually sparse, ie, contains many zeros. To account for the zero-inflated structure, we propose a multivariate zero-inflated logistic-normal (MZILN) distribution for the data. Let $Z$ be a $(K+1)$-dimensional vector containing 1's and 0's with the $k$th element $Z_k = 1/0$ to indicate the $k$th taxon being positive/zero. It is straightforward to see that the observed vector $Y$ can be expressed in terms of $Y^*$ and $Z$:

$$Y = \left( \frac{Y_1^* Z_1}{\sum_k Y_k^* Z_k}, \ldots, \frac{Y_{K+1}^* Z_{K+1}}{\sum_k Y_k^* Z_k} \right)^T.$$

Under the assumption that $Y^*$ follows a LN distribution, naturally $Y$ will follow a MZILN distribution with two parts: discrete part that governs the probabilities of elements in $Z$ being 0 or 1, and the continuous part that provides the conditional distribution function for the log-ratio transformation of observed non-zero RA's.

Let $p_{k_1,\ldots,k_L}$ denote the probability that the subset elements $Z_{k_1}, \ldots, Z_{k_L}$ in $Z$ are 1 and all other elements in $Z$ are zero. The distribution for discrete part can be written as:

$$P(Z_1 = 1, Z_2 = 0, \ldots, Z_{K+1} = 0) = p_1,$$

$$P(Z_1 = 0, Z_2 = 1, \ldots, Z_{K+1} = 0) = p_2,$$

$$\ldots$$

$$P(Z_1 = 0, \ldots, Z_K = 0, Z_{K+1} = 1) = p_{K+1},$$

$$\ldots$$

$$P(Z_1 = 0, \ldots, Z_{k_1-1} = 0, Z_{k_1} = 1, Z_{k_1+1} = 0, \ldots, Z_{k_L} = 1, Z_{k_L+1} = 0, \ldots, Z_{K+1} = 0) = p_{k_1,\ldots,k_L},$$

$$\ldots$$

$$P(Z_1 = 1, \ldots, Z_{K+1} = 1) = p_{1,\ldots,K+1},$$

and

$$\sum_{\substack{1 \leq k_1 < \cdots < k_L \leq K+1 \\ 1 \leq L \leq K+1}} p_{k_1,\ldots,k_L} = 1.$$

There are $(2^{K+1} - 2)$ parameters (i.e., $p_{k_1,\ldots,k_L}$'s) in the discrete part. This is essentially a $(K+1)$-dimensional multivariate Bernoulli distribution [36, 37] conditional on at least one element being 1.

The vector $Z$ is similar to a missing indicator vector except that here it indicates whether the observed RA is positive or zero. For any taxon, say the $k$th taxon, there could be two reasons for $Z_k = 0$: a) the taxon is truly absent and b) the taxon is not truly absent, but somehow it does not have any sequencing reads. It can be shown that $Y_k > 0$ is equivalent to $Z_k = 1$ for all $k = 1, \ldots, K+1$ (See details in Section S5 of the supplemental material). So the distribution of the discrete part can be also rewritten in terms of $Y$ as follows:

$$P(Y_1 > 0, Y_2 = 0, \ldots, Y_{K+1} = 0) = p_1,$$

$$\ldots$$

$$P(Y_1 = 0, \ldots, Y_K = 0, Y_{K+1} > 0) = p_{K+1},$$

$$\ldots$$

$$P(Y_1 = 0, \ldots, Y_{k_1-1} = 0, Y_{k_1} > 0, Y_{k_1+1} = 0, \ldots, Y_{k_L} > 0, Y_{k_L+1} = 0, \ldots, Y_{K+1} = 0) = p_{k_1,\ldots,k_L},$$

$$\ldots$$

$$P(Y_1 > 0, \ldots, Y_{K+1} > 0) = p_{1,\ldots,K+1}.$$

Conditional on the subset $Y_{k_1}, \ldots, Y_{k_L}$ being non-zero and all other elements of $Y$ being zero, the observed RA vector $Y = \left(0, \ldots, 0, \frac{Y_{k_1}^*}{\sum_{l=1}^L Y_{k_l}^*}, 0, \ldots, 0, \frac{Y_{k_L}^*}{\sum_{l=1}^L Y_{k_l}^*}, 0, \ldots, 0\right)^T$. We know that the subcomposition of the non-zero RA's $\left(\frac{Y_{k_1}^*}{\sum_{l=1}^L Y_{k_l}^*}, \frac{Y_{k_2}^*}{\sum_{l=1}^L Y_{k_l}^*}, \ldots, \frac{Y_{k_L}^*}{\sum_{l=1}^L Y_{k_l}^*}\right)^T$ follows a LN distribution from the previous section. Thus, the density function of continuous part is given by

$$f(y) = \begin{cases} p_{k_1,\ldots,k_L} g(u_{k_1,\ldots,k_L}), & y = (0,\ldots,0, y_{k_1}, 0, \ldots, 0, y_{k_L}, 0, \ldots, 0)^T, \\ \vdots \\ p_{1,\ldots,K+1} h(u), & y = (y_1, \ldots, y_{K+1})^T, \end{cases}$$

where $y$, $u$ and $u_{k_1,\ldots,k_L}$ are the realizations of the random vectors $Y$, $U$ and $U_{k_1,\ldots,k_L}$ respectively, and $g(u_{k_1,\ldots,k_L})$ and $h(u)$ are the density functions of the two multivariate normal distributions $N(A\mu, A\Sigma A^T)$ and $N(\mu, \Sigma)$ respectively. The density function $f(y)$ involves discrete probability masses $p_{k_1,\ldots,k_L}$'s because the continuous part of MZILN is essentially a distribution conditional on the subset $Y_{k_1}, \ldots, Y_{k_L}$ being non-zero.

In summary, the MZILN distribution is fully determined by these parameters: the mean vector $\mu$, the variance matrix $\Sigma$, and the discrete probability masses $p_{k_1,\ldots,k_L}$, $1 \leq k_1 < \cdots < k_L \leq K+1$, $1 \leq L \leq K+1$.

## 2.3 Regression model

Let $x_i$ be the $Q$ by 1 vector of covariates and $\mu^i$ denote the $K$-dimensional mean vector of $U$ for the $i$th subject. The regression model for the mean is

$$\mu^i = X_i\beta, \quad (2)$$

where $X_i = I_K \otimes (1, x_i^T)$, Kronecker product, is a $K \times M$ matrix of covariates where $M = K(Q+1)$ and $\beta = (\beta_{01}^T, \ldots, \beta_{0K}^T)^T$ is a $M$-dimensional vector of regression coefficient parameters. Here $\beta_{0k}$ is the $(Q+1)$-dimensional vector of parameters associated with the $k$th element of the mean vector $\mu^i$. If we write $\beta = (\beta_1, \ldots, \beta_p)^T$, then $\beta_{0k} = (\beta_{(k-1)(Q+1)+1}, \beta_{(k-1)(Q+1)+2}, \ldots, \beta_{k(Q+1)})^T$, $k = 1, \ldots, K$. We can also extract the $K$-dimensional vector of parameters associated with the $q$th covariate: $\beta_{q0} = (\beta_{q+1}, \beta_{q+1+(Q+1)}, \ldots, \beta_{q+1+(K-1)(Q+1)})^T$, $q = 0, 1, \ldots, Q$. Vector $\beta_{q0}$ becomes the intercept vector when $q = 0$. Let $\beta_{q0}^k$ denote the $k$th, $k = 1, \ldots, K$, element of $\beta_{q0}$. A straightforward interpretation for $\beta_{q0}^k$ is that it denotes the amount of change in RA of the $k$th taxa on log scale given one unit increase in the $q$th covariate, controlling for other covariates and the $(K+1)$th taxa.

The overall perturbation can be also quantified in terms of the parameters with the assistance from a perturbation operator [26, 34, 38]. The vector

$$\left(\frac{\exp(\beta_{00}^1)}{1 + \sum_{k=1}^K \exp(\beta_{00}^k)}, \ldots, \frac{\exp(\beta_{00}^K)}{1 + \sum_{k=1}^K \exp(\beta_{00}^k)}, \frac{1}{1 + \sum_{k=1}^K \exp(\beta_{00}^k)}\right)$$

represents the baseline microbiome composition without disturbance from any of the covariates. The vector

$$\left(\frac{\exp(\beta_{q0}^1)}{1 + \sum_{k=1}^K \exp(\beta_{q0}^k)}, \ldots, \frac{\exp(\beta_{q0}^K)}{1 + \sum_{k=1}^K \exp(\beta_{q0}^k)}, \frac{1}{1 + \sum_{k=1}^K \exp(\beta_{q0}^k)}\right)$$

measures the shift in composition from baseline by one unit change in the $q$th covariate. The association of the covariate with the $k$th taxon is positive if $k$th element greater $\frac{1}{K+1}$ and negative if less. The magnitude of overall disturbance in microbiome composition induced by one unit change in the $q$th covariate is measured by $\sqrt{\beta_{q0}^T (I_K + 1_K 1_K^T)^{-1} \beta_{q0}}$ where $I_K$ is the $K \times K$ identity matrix and $1_K$ is the $K$-dimensional vector of 1's.

We can also model the associations between covariates $x_i$ and the discrete part of the MZILN distribution by allowing the parameters $p_{k_1,\ldots,k_L}$, $1 \le k_1 < \cdots < k_L \le K+1$, $1 \le L \le K+1$ to depend on the covariates $x_i$. The parameters describing the associations between covariates $x_i$ and the parameters $p_{k_1,\ldots,k_L}$ can be treated as nuisance parameters. So we will leave out that part. More details can be found in Section S1 of the supplemental material.

## 2.4 Estimation: estimating equation approach based on likelihood function

In this paper, we are interested in estimating the parameter vector $\beta$ that characterize the associations between the covariates and log-ratio transformed microbiome taxa RA. We will propose an estimating equation approach for the estimation based on log-likelihood function. Let $A_i$, $U^i$, $U^i_{k_1,\ldots,k_L}$ and $p^i_{k_1,\ldots,k_L}$ denote the counterparts of $A$, $U$, $U_{k_1,\ldots,k_L}$ and $p_{k_1,\ldots,k_L}$ for the $i$th subject. We divide subjects into two groups based on the availability of taxa RA data: 1) subjects with only one non-zero RA and 2) subjects with two or more non-zero RA's. The full log-likelihood function is just the summation of the log-likelihood contributions from those two groups.

For the first group, the log-likelihood contribution comes only from the discrete part, and thus it can be written as $log(p^i_{k_1})$ where the sup script $i$ is subject index and $k_1$ denote the taxon with non-zero RA.

For the second group, the log-likelihood contribution comes from both the discrete part and the continuous part. Without loss of generality, let $Y^i_{k_1},\ldots,Y^i_{k_L}$ be the non-zero RA's for the $i$th subject in this group. Under the regression model, the vector $U^i_{k_1,\ldots,k_L}$ follows the normal distribution with mean $A_i X_i \beta$ and variance matrix $A_i \Sigma A_i^T$. Notice that when all RA's are non-zero, $Y^i_{k_1},\ldots,Y^i_{k_L}$ are simply the RA's of all the taxa and $A_i$ becomes the $K \times K$ identity matrix. Thus the log-likelihood contribution from this subject is

$$log(p^i_{k_1,\ldots,k_L}) + 0.5 \log |A_i \Sigma A_i^T|^{-1} - 0.5 (U^i_{k_1,\ldots,k_L} - A_i X_i \beta)^T (A_i \Sigma A_i^T)^{-1} (U^i_{k_1,\ldots,k_L} - A_i X_i \beta) + \text{constant}.$$

Summing together the log-likelihood contributions from all subjects, we can write the complete log-likelihood function as:

$$\sum_i \log(p^i_{k_1,\ldots,k_L}) + \sum_i 0.5 \log |A_i \Sigma A_i^T|^{-1} - 0.5 \sum_i (U^i_{k_1,\ldots,k_L} - A_i X_i \beta)^T (A_i \Sigma A_i^T)^{-1} (U^i_{k_1,\ldots,k_L} - A_i X_i \beta) + \text{constant}.$$

Notice the terms involving parameters $\beta$ and $\Sigma$ do not depend on $p^i_{k_1,\ldots,k_L}$'s, and thus they can be maximized separately to obtain MLEs of the parameters $\beta$ and $\Sigma$ by treating $p^i_{k_1,\ldots,k_L}$'s as nuisance parameters.

Let $\Omega_i = (A_i \Sigma A_i^T)^{-1}$, the new objective function involving only $\beta$ and $\Sigma$ can be written as:

$$l(\beta, \Sigma) = 0.5 \sum_i \log \Omega_i - 0.5 \sum_i (\tilde{U}_i - \tilde{X}_i \beta)^T (\tilde{U}_i - \tilde{X}_i \beta),$$

where $\tilde{U}_i = \Omega_i^{1/2} U^i_{k_1,\ldots,k_L}$ and $\tilde{X}_i = \Omega_i^{1/2} A_i X_i$. The parameters $\beta$ and $\Sigma$ can be estimated by setting the partial derivatives of objective function to 0. The equation with respect to $\beta$ is:

$$\sum_i (\tilde{X}_i)^T (\tilde{U}_i - \tilde{X}_i \beta) = 0, \quad (3).$$

There is another much more complicated equation for $\Sigma$ as well. When the dimension of $\Sigma$ is not high (e.g., the number of taxa less than sample size), we can solve these equations to obtain maximum likelihood estimators for both $\beta$ and $\Sigma$.

For high dimensional cases, however, it is computationally challenging to estimate Σ and the MLE of Σ is usually not stable [39]. Fortunately, equation (3) is an estimating equation because the expectation of left-hand side is equal to 0, and thus equation (3) will produce consistent estimator of $\beta$ for any fixed (could be mis-specified) covariance matrix Σ [29, 40]. For simplicity and speed, we choose Σ to be the identity matrix. This is similar to the independence correlation structure under a GEE setting. Furthermore, the solution of equation (3) minimizes the sum of square error $\sum(\widetilde{U}_i - \widetilde{X}_i\beta)^T(\widetilde{U}_i - \widetilde{X}_i\beta)$, and therefore the estimator of $\beta$ becomes ordinary least-square (OLS) estimator given by

$$\hat{\beta} = (\widetilde{X}^T\widetilde{X})^{-1}\widetilde{X}^T\widetilde{U},$$

where $\widetilde{U} = \begin{pmatrix} \widetilde{U}_1 \\ \vdots \\ \widetilde{U}_N \end{pmatrix}$ and $\widetilde{X} = \begin{pmatrix} \widetilde{X}_1 \\ \vdots \\ \widetilde{X}_N \end{pmatrix}$. Due to the high dimensionality of $\beta$, a sparse estimate is desired to have easy and straightforward interpretation. Regularization approaches have been well established for OLS estimator such as LASSO [30], adaptive LASSO [41], Elastic Net [42], SCAD [31] and MCP [32]. While all the regularization approaches can be used, we will illustrate our approach with the MCP method where the tuning parameter is selected by minimizing the mean square error of a 10-fold cross validation. Our simulations showed that MCP gave better performance in identifying the true taxa.

## 3 Simulation

### 3.1 Simulation with low dimensionality: K<N

To examine the asymptotic properties of the estimators under low dimensional settings, three hundred data sets were randomly generated with each data set having 1000 subjects and 20 taxa (K=19). A 20-dimensional multivariate Bernoulli distribution was used to generate the discrete part where the marginal Bernoulli distributions were assumed to be independent. All the Bernoulli distributions have the same probability of 0.5 to be zero. A single covariate was generated from the standard normal distribution for the regression model. All intercept parameters in $\beta_{00}$ were set to be -0.1 and all coefficients parameters in $\beta_{10}$ were set to be 0.8. The variance matrix Σ is set to have diagonal elements being 1 and off-diagonal elements being 0.3. This corresponds to an exchangeable correlation structure with $\rho = 0.3$. We calculated the average bias (Ave.Bias) of point estimators. The average percent of bias (Ave.Percent.Bias) and the average empirical coverage probabilities (Ave.CP) of the 95% confidence intervals (CI) were obtained as well. Results (Table 1) show that the estimator is virtually unbiased and CP is reasonably close to 95%.

**Table 1. Simulation results for low dimensional case.** Ave.Bias is the average bias of the estimates for the 19 parameters; Ave.Percent.Bias is the average bias as the percentage of the true value; Ave.CP is the average empirical CP of the 95% CI for the parameters.

| Parameter | True | Ave.Bias | Ave.Percent.Bias (%) | Ave.CP (%) |
|---|---|---|---|---|
| $\beta_{00}$ | -0.1 | 0.0003 | 2.00 | 94.8 |
| $\beta_{10}$ | 0.8 | 0.0004 | 0.17 | 94.7 |
| SD | 1 | -0.003 | 0.33 | 94.5 |
| $\rho$ | 0.3 | -0.0004 | 0.15 | 94.4 |

### 3.2 Simulation with high dimensionality: K>N

We carried out simulation studies to evaluate the performance of our proposed model for high dimensional cases under a number of settings. First, we assessed the impact of over-dispersion on model performance. Under the MZILN model, 100 data sets were randomly generated with each data set having $N = 300$ subjects, $K + 1 = 400$ taxa (e.g., genera) and $Q = 40$ covariates. There were $K \times (Q + 1) = 16359$ regression coefficients under this setting. The covariates were generated using a 40-dimensional multivariate normal distribution with mean 0 and a polynomial decay variance matrix with the $ij$th element equal to $\rho_X^{|i-j|}, i,j = 1, \ldots, 40$ where $\rho_X = 0.5$. We assumed that only 4 covariates were truly associated with the microbiome community and each of the 4 covariates was associated with 9 log-ratio transformed taxa. That means the 16359-dimensional $\beta$ vector had only 36 non-zero elements which were generated from a uniform distribution over the interval $[-3, -1) \cup (1, 3]$. To mimic the non-zero RA proportion in real data, the probability of having non-zero RA is set to be 0.54 for each taxon. We set the $ij$th element of outcome variance matrix $\Sigma$ to be $\sigma^2 \rho^{|i-j|}, i,j = 1, \ldots, 399$, where $\rho = 0.5$ and $\sigma$ was chosen to control the average signal-to-noise ratio (SNR). The SNR can be translated into over-dispersion according to their inverse relationship [26]. A high/low SNR indicates a low/high over-dispersion and there is no over-dispersion if SNR is infinity (ie, $\sigma = 0$). We tested three scenarios with high, moderate and low over-dispersion by setting SNR equal to 1.5, 4.5 and 7.5 respectively. We evaluated the model performance by the three measures: recall=TP/(TP+FN), precision=TP/(TP+FP) and F1=2*recall*precision/(recall+precision), where TP, FN and FP denote true positives, false negatives and false positives, respectively, and F1 is an overall measure weighting the precision and recall equally. We compared different regularization approaches including LASSO [30], adaptive LASSO [41], Elastic Net [42], SCAD [31] and MCP [32], and MCP gave the best model performances (See Section S2 in supplemental material). Thus, we present simulation results with MCP employed as the regularization approach. The results (Figure 1A) showed that the model performs better as over-dispersion decreases. The model can accommodate over-dispersion very well as all the performance measures were good across all the three scenarios. The high recall rates indicate that the model is powerful in terms of picking up the non-zero coefficients. The good precision rates indicate low false positive rates. The F1 score had a similar pattern as recall and precision rates.

Second, we examined the robustness of our approach with respect to misspecification of the outcome correlations. Three cases with weak, moderate and strong correlations were tested where $\rho$ was set to be 0.2, 0.5 and 0.8 respectively. Data were generated with SNR=4.5 and other parameter settings were the same as previously described for testing the effects of over-dispersion. Results (Figure 1B) showed that the model is insensitive to correlation misspecification as the performance measures remain relatively stable for all three situations. The recall, precision and F1 measures are not only stable, but also having high values across the three cases which again marks the good model performance.

As suggested by one of the reviewers, we also examined the robustness with respect to misspecification of the distribution on top of the misspecification of correlation. Correctly specified regression equation is $U^i = X_i\beta + \varepsilon^i$, where $i$ is subject index and $\varepsilon^i$ have the normal distribution $N(0, \Sigma)$. We add a perturbation to the residual so that the distribution is mis-specified: $U^i = X_i\beta + (1-\gamma)\varepsilon^i + \gamma\sigma(\delta^i - 1)$ where $0 \leq \gamma \leq 1$ and $\delta^i$ is a random vector with each element following the chi-square distribution with 1 degrees of freedom. The parameter $\sigma$ is to adjust signal-to-noise ratio. The two random vectors $\varepsilon^i$ and $\delta^i$ are independent. Notice that $\gamma$ quantifies the degree to which the model is mis-specified. $\gamma = 0$ corresponds to the correctly specified distribution, $\gamma = 1$ corresponds to a completely misspecified distribution, and $0 < \gamma < 1$ corresponds to a partially mis-specified distribution. In this set of simulations, $\rho_X = 0.85$, $\rho = 0.5$, SNR=4.5, data sparsity is set at 0.54 and the non-zero regression coefficients were generated from a uniform distribution over the interval $[-7, -4) \cup (4, 7]$. All other settings are the same as described at the beginning of this section. The results (See Section S3 in the supplemental material) showed that the recall rate is fairly robust to this misspecification. The precision dropped a little bit, but it remains stable as $\gamma$ increases. F1 has a similar pattern as precision.

Third, we evaluated the model performance under different data sparsity levels. Previously, each taxon was set to have $p = 54\%$ non-zero RA. Here we simulated two more situations: one with a low sparsity level ($p = 0.2$) and the other with a high sparsity level ($p = 0.8$). SNR and $\rho$ were fixed at 4.5 and 0.5 respectively and all other parameters were the same as described at the beginning of this section. Results showed (Figure 1C) that our approach can handle all the three scenarios ranging from high data sparsity to low data sparsity. Recall rates were high across the three sparsity levels and, similar to earlier simulation results, good precision rates and F1 scores were observed as well. The high data sparsity level did not have a strong negative impact on the performance measures.

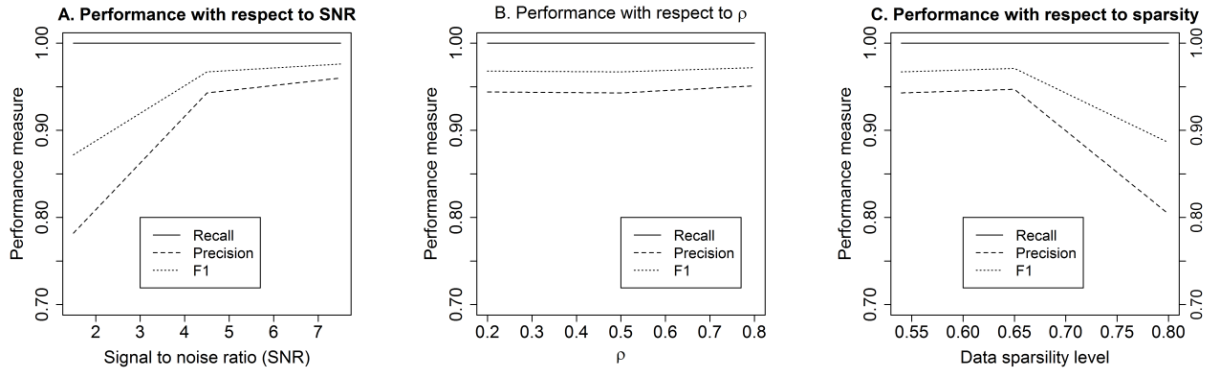

Fig. 1: Model performance measures as a function of the SNR (in panel A), the correlation (in panel B) and the data sparsity level (in panel C).

We performed two additional sets of simulations where we randomly chose different reference taxon to check the robustness of our model for different reference taxon. In these simulations, $\rho_X = 0.85$, $\rho = 0.5$, SNR=4.5, data sparsity is set at 0.54 and the non-zero regression coefficients were generated from a uniform distribution over the interval $[-7, -4) \cup (4, 7]$. All other settings are the same as described at the beginning of this section. The results (See Section S4 in supplemental material) showed that the recall rate had good robustness compared with the case with the true reference taxon (i.e., the reference taxon used in the data generation). Precision rate and F1 score dropped a little bit, but they remained stable across the two cases with randomly selected reference taxon.

### 3.3 Comparisons with other methods

We also compared our approach with established existing approaches: the sparse Dirichlet-multinomial (DM) regression [25], kernel-penalized regression (KPR) [22], zero-inflated beta (ZIB) regression [27] and the nonparametric correlation: Spearman (SP) correlation test. KPR, ZIB and SP employ the false discovery rate (FDR) control for correcting multiple comparisons. KPR employ a significance test [43] to generate p values after penalized estimates are obtained. ZIB and SP test each covariate-taxon association one by one and selected the pairs based on the FDR control. We set FDR=0.05 in the simulation. The comparison was carried out under three SNR levels (1.5, 4.5, 7.5) and three data sparsity levels (0.54, 0.65, 0.8). The data sparsity level was set at 0.54 when studying different SNR levels. The SNR was set at 4.5 when studying different data sparsity levels. Other simulation settings are the same as described at the beginning of this section except that the value of $\rho_X$ was changed to 0.85 and the non-zero elements of the $\beta$ vector were generated from a uniform distribution over the interval $[-7, -4) \cup (4, 7]$.

The results (Fig. 2) showed that our approach outperforms all other approaches by a wide marge in terms of recall rate and F1. The precision rate of our approach is also superior for most of the time except when data sparsity level is high where ZIB has higher precision rate (Fig. 2E). This is probably due to the smaller average model size:18.2 for ZIB. A downside of the ZIB approach [27] is that it does not provide effect size estimates, and consequently it is unknown whether an identified taxon is positively or negatively associated with a covariate.

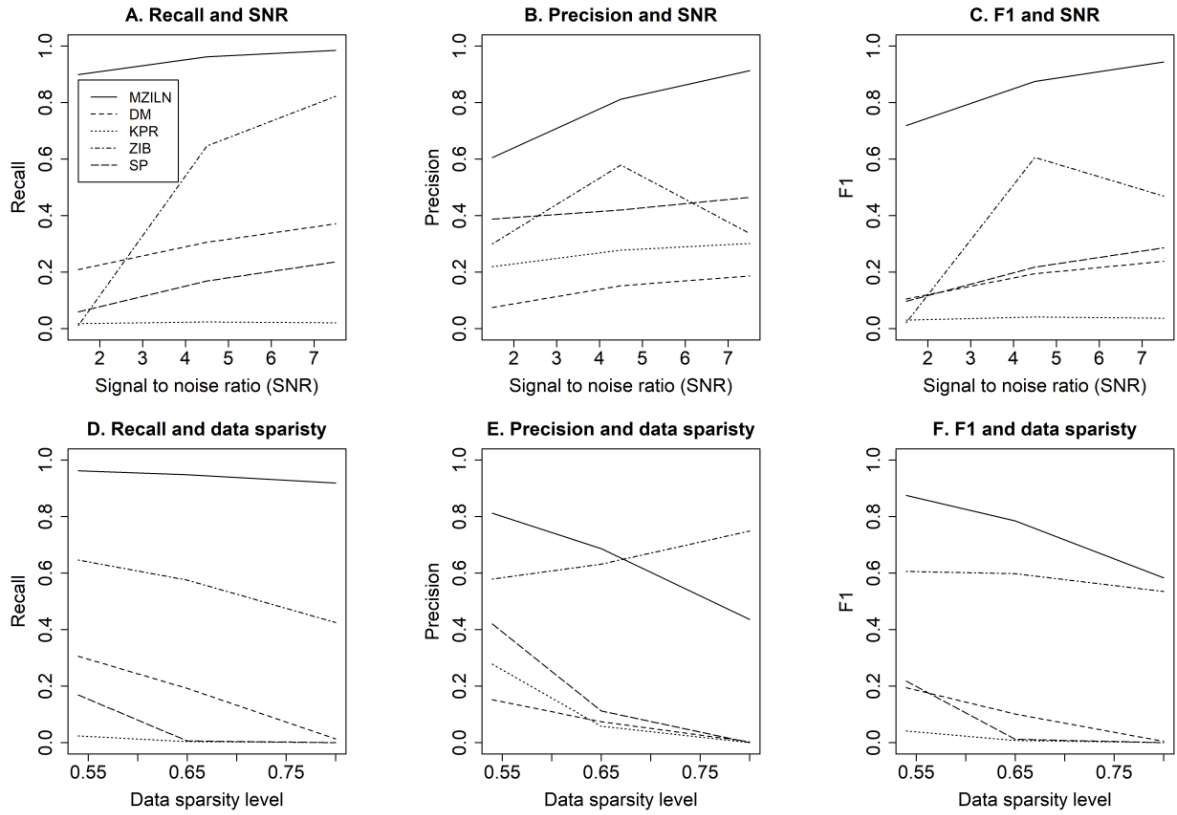

Fig 2: Performance comparison of our approach (MZILN) with the sparse Dirichlet-multinomial regression (DM), kernel-penalized regression (KPR), zero-inflated beta regression (ZIB) and Spearman's correlation test (SP). FDR was set at 0.05 for KPR, ZIB and SP.

## 4 Application in the New Hampshire Birth Cohort Study

New Hampshire Birth Cohort Study (NHBCS) is a large ongoing molecular epidemiological cohort study to evaluate the health impacts of environmental exposures with a focus on arsenic in pregnant women and their children in rural New England [44]. The study began enrollment of pregnant women at about 24 to 28 weeks prenatal appointments at study clinics and follow up both mothers and babies after birth. Madan and Hoen et al. [45] studied the associations of delivery mode and feeding method with infant intestinal microbiome composition at approximately 6 weeks of life in a subset of approximately 100 full-term babies from the NHBCS. Participants provided infant stool samples collected at six weeks postpartum. Delivery mode (cesarean vs. vaginal delivery) was abstracted from maternal delivery records. About 30% babies were operatively delivered by Cesarean section and the rest were vaginally delivered. Feeding method was determined by interval telephone interviews about infant diet from birth until the time of stool collection. Feeding type was grouped into three categories: breast fed, formula fed and mixed fed with approximately 70%, 6% and 25% babies in these categories respectively. DNA was extracted from the stool samples using the Zymo DNA extraction kit (Zymo Research). Illumina tag sequencing of the 16S rRNA gene V4-V5 hypervariable region was performed at the Marine Biological Laboratories (MBL) in Woods Hole, MA with established methods [46, 47]. Using QIIME version 1.9.1 (74), open reference operational taxonomic units (OTUs) were formed from the sequences with the uclust algorithm at 97% similarity (75). PyNAST alignment (76) with Greengenes core reference (77, 78) on the representative sequence for each OTU was used to build the OTU-table and assign taxonomy (78, 79). A phylogenetic tree was constructed using the FastTree method (80). 16S sequencing generated a total of 14,362,739 (mean: 140,811, range: 27,897 – 260,579) bacterial DNA reads, of which, 8,210,402 (mean: 80,494, range: 12,244 - 178,802) passed quality filters and formed 8612 OTUs that were assigned to 253 bacterial genera.

We reanalyzed the data using our method to identify individual taxa that are differently abundant across delivery modes and feeding types. In the statistical data analysis, 12 genera were removed because they had no sequencing reads on those subjects who had information on both delivery mode and feeding type. We have $K = 241$ in the analysis since there were 241 genera in the data, and thus the vector $U$ is an 240-dimensional vector. *Akkermansia* was set as the reference genus at random. There were two covariates in the regression model: delivery mode and feeding type. There were 2*240=480 regression coefficients in the model. We coded delivery mode as a binary independent variable (0=cesarean, 1=vaginal delivery). Due to the small number of formula fed babies, and because in the previous analyses we identified microbiome patterns in mixed fed babies were more similar to formula fed than exclusively breastfed babies, we lumped formula fed and mixed fed babies together such that feeding type was also a binary variable (0=breast fed, 1=formula or mixed fed). Other covariates can be easily added to the model if necessary. MCP was used as the regularization approach in our analysis.

There were 28 genera selected for the association with delivery mode (Table 2), of which 17 genera had positive associations and 11 genera had negative associations. Compared with Madan and Hoen et al. [45] that only found 5 genera in association with delivery mode, although we missed two of their genera (*Pectobacterium* and *Rothia*), our approach found 23 more genera including *Bifidobacterium, Clostridium* and *Streptococcus* which are known to have important impact on children's health [48-57] . There were 23 genera selected for association with feeding type (Table 3), of which 9 genera had positive associations and 14 genera had negative associations. Madan and Hoen et al. [45] found feeding type associated with only one genus (*Lactococcus*) which was also selected by our method, and in addition, we identified 22 more genera including *Bacteroides*, *Bifidobacterium*, *Blautia* and *Enterococcus* that have been linked to infant's health in the literature [48, 57-64].

As a sensitivity analysis, we randomly chosen a different reference genus (*Anoxybacillus*) and reran our approach on the real data set, the selected genera are generally consistent (See Table 1A in the Appendix) especially for those with stronger associations. For example, the top 8 genera positively associated with feeding type are the same. The top 8 genera negatively associated with feeding type are also the same. For the genera positively associated with delivery mode, almost all genera are the same except 1 (out of 16) genus identified by reference genus *Akkermansia* was not identified by reference genus *Anoxybacillus* and 3 (out of 18) genera identified by reference genus *Anoxybacillus* were not identified by reference genus *Akkermansia*. For genera negatively associated with delivery mode, the top 4 genera identified by reference genus *Anoxybacillus* are among the top 6 genera identified by reference genus *Akkermansia*.

As a comparison, we also analyzed the data using DM, ZIB and SP. We also applied Wilcoxon rank sum test which generated nearly identical results as the SP approach, thus Wilcoxon test results were not presented. We did not include KPR in this comparison because KPR is not developed for testing the associations of binary variables with microbiome. FDR was set at 0.05 for ZIN and SP. Consistent with the simulation results, ZIB and SP found less genera than MZILN as shown in Tables 2 and 3. DM selected more taxa than expected and had good overlap with MZILN.

Table 2. Genera identified to be associated with delivery mode. Black and green indicate positive and negative associations respectively. Red indicates that the direction of the identified association is unknown. The genera are sorted by association strength (measured by magnitude of estimated effect size or p value) from strongest to weakest in each category.

| MZILN | DM | ZIB | Spearman |
|---|---|---|---|
| *Bacteroides* | *Bacteroides* | *Bacteroides* | *Bacteroides* |
| *Phascolarctobacterium* | *Parabacteroides* | | *Sutterella* |
| *Parabacteroides* | *Sutterella* | | *Parabacteroides* |
| *Eubacterium* | *Collinsella* | | *Bilophila* |
| *Megamonas* | *Bifidobacterium* | | |
| *Collinsella* | *Phascolarctobacterium* | | *Clostridium* |
| *Bifidobacterium* | *Prevotella* | | *Veillonella* |
| *Prevotella* | *Bilophila* | | *Serratia* |
| *Ruminococcus* | *Escherichia* | | *Staphylococcus* |
| *Faecalibacterium* | *Eggerthella* | | *Streptococcus* |
| *Escherichia* | | | |
| *Corynebacterium* | *Clostridium* | | |
| *Lactobacillus* | *Streptococcus* | | |
| *Chryseobacterium* | *Veillonella* | | |
| *Coprobacillus* | *Serratia* | | |
| | *Enterococcus* | | |
| *Clostridium* | *Staphylococcus* | | |
| *Veillonella* | *Citrobacter* | | |
| *Propionibacterium* | *Finegoldia* | | |
| *Serratia* | *Eubacterium* | | |
| *Atopobium* | *Corynebacterium* | | |
| *Haemophilus* | *Actinomyces* | | |
| *Actinomyces* | *Atopobium* | | |
| *Dorea* | *Chryseobacterium* | | |
| *Staphylococcus* | *Propionibacterium* | | |
| *Finegoldia* | *Haemophilus* | | |
| *Streptococcus* | *Erwinia* | | |

| MZILN | DM | ZIB | Spearman |
|---|---|---|---|
| *Eubacterium* | *Enterococcus* | <span style="color:red">*Enterococcus*</span> | *Lactococcus* |
| *Enterococcus* | *Lactococcus* | <span style="color:red">*Staphylococcus*</span> | *Enterococcus* |
| *Oscillospira* | *Eubacterium* | | *Oscillospira* |
| *Ruminococcus* | *Oscillospira* | | *Eubacterium* |
| *Lactococcus* | *Granulicatella* | | *Kocuria* |
| *Blautia* | *Peptoniphilus* | | *Granulicatella* |
| *Dorea* | *Anaerococcus* | | |
| *Collinsella* | *Finegoldia* | | <span style="color:green">*Haemophilus*</span> |
| | *Blautia* | | <span style="color:green">*Staphylococcus*</span> |
| <span style="color:green">*Haemophilus*</span> | *Streptococcus* | | |
| <span style="color:green">*Staphylococcus*</span> | *Ruminococcus* | | |
| <span style="color:green">*Serratia*</span> | *Eggerthella* | | |
| <span style="color:green">*Propionibacterium*</span> | *Dorea* | | |
| <span style="color:green">*Citrobacter*</span> | *Kocuria* | | |
| <span style="color:green">*Corynebacterium*</span> | | | |
| <span style="color:green">*Bifidobacterium*</span> | <span style="color:green">*Haemophilus*</span> | | |
| <span style="color:green">*Escherichia*</span> | <span style="color:green">*Staphylococcus*</span> | | |
| <span style="color:green">*Rothia*</span> | <span style="color:green">*Limnohabitans*</span> | | |
| <span style="color:green">*Peptoniphilus*</span> | <span style="color:green">*Comamonas*</span> | | |
| <span style="color:green">*Clostridium*</span> | <span style="color:green">*Corynebacterium*</span> | | |
| <span style="color:green">*Acinetobacter*</span> | <span style="color:green">*Propionibacterium*</span> | | |
| <span style="color:green">*Bacteroides*</span> | <span style="color:green">*Phenylobacterium*</span> | | |
| <span style="color:green">*Pseudomonas*</span> | <span style="color:green">*Bifidobacterium*</span> | | |
| | <span style="color:green">*Serratia*</span> | | |
| | <span style="color:green">*Klebsiella*</span> | | |

Table 3. Genera identified to be associated with feeding type. Black and green indicate positive and negative associations respectively. Red indicates that the direction of the identified association is unknown. The genera are sorted by association strength (measured by magnitude of estimated effect size or p value) from strongest to weakest in each category.

## 5 Discussion

This paper proposed an innovative MZILN model for analyzing microbiome RA in relation to health risk factors. The approach is essentially a two-part model with the discrete part to handle excessive number of zeros commonly seen in microbiome sequencing data and the logistic-normal part to address the compositional structure of microbiome RA data. Standard regularization procedures such as LASSO, SCAD and MCP can be easily incorporated into this approach to obtain sparse estimations of high-dimensional regression parameters to avoid overfitting of the model. By borrowing the strength of estimating equations, the proposed approach can accommodate complex inter-taxa correlation structure induced by the phylogenetic hierarchical structure and the compositional data structure. Our simulation study has demonstrated the performance of our approach in comparison with existing methods. Our approach can be applied to RA of OTU, amplicon sequence variant and other RA data as well although the description in this paper has been focusing on analyzing taxa RA. R program is available upon request to implement the method. We are also working on building an R package.

Compared with the miLineage approach [24], an immediate advantage of our method is the flexibility to handle high-dimensional microbial taxa data (ie, number of taxa bigger than sample size) with regularization approaches whereas their approach has to analyze lineages to have a solution in such cases. Depending on what is needed in practice, our model can produce sparse estimates with individual $\ell_1$ penalties as well as group $\ell_1$ penalties. Our handling of high dimensionality is different than those methods that treat microbiome data as covariates instead of outcome variables [21, 23] where standard regularization approaches cannot be directly applied due to the compositional structure of the covariates. Penalized likelihood estimation methods have also been developed to analyze high-dimensional microbiome absolute abundance count data in relation to other covariates such as micronutrients [25, 26], but they are not as flexible as our method in terms of employing the penalization terms. Our estimator has a very simple form: ordinary least square (OLS) estimator, and thus naturally allows for all standard regularization approaches that can be applied for OLS estimators. A downside of our proposed approach is that we did not consider the zero-part in the estimation by treating the zero-part parameters as nuisance parameters. This is equivalent to a conditional regression where only the positive data points contribute to the estimation. This may cause efficiency loss in the estimation process when microbiome data is extremely sparse. However, even with sparse data, the overall performance of our approach still is still better than other approaches according to the comparison in the simulation study.

Compared with many existing methods developed to analyze RA data, one of the nice properties of our method is that we do not impute zero sequencing counts with a pseudo count (eg, 0.5) or impute zero proportion with an arbitrary small proportion. When dealing with RA data, log-ratio transformation is often used to address the compositional data structure. However, log-ratio transformation can only be applied to non-zero RA, and thus imputation for zero RA is a commonly used technique in the literature which could distort the data and consequently distort the estimated associations. Our method does not need to impute zero RA's by constructing the MZILN distribution that can appropriately handle the zero-inflated data structure.

Our approach allows for a very flexible inter-taxa correlation structure. There are two main drivers for the inter-taxa correlation: the inherent compositional data structure and the hierarchical phylogenetic tree structure. Compositional data structure induces negative correlations between taxa because all taxa RA sum to 1 and thus one RA increase is accompanied by the decrease of another RA. The phylogenetic tree structure reflects evolutionary relationships among microbes based upon similarities and differences in their genetic characteristics. It does not necessarily induce negative inter-taxa correlations. Depending on the functional relationships of microbes, this hierarchical tree structure could generate positive or negative inter-taxa correlations. The compositional data structure and the hierarchical phylogenetic tree structure are compounded in the data and can generate complicated inter-taxa correlations. Our MZILN method adequately handle the complex correlation structure by utilizing powerful estimation tools from estimating equations approaches.

Although normal distribution is assumed for the log-ratio transformation of the data, this assumption can be largely relaxed in practice since estimating equation (3) does not rely on the normal distribution assumption as long as the mean of the left side of equation (3) is 0. The robustness to mis-specification of distribution was demonstrated with a simulation. This allows real data analysis to address a much broader range of distributions, and thus it make the model a very useful tool for researchers to study associations of microbiome with other variables of interest.

The proposed approach needs to select a reference taxon because of the definition of the logistic-normal distribution. Simulation study showed that results are reasonably stable across randomly selected reference taxa. In the real data application, we also saw good consistent results across two randomly selected reference taxa although there are some differences. Our method is flexible in choosing a reference taxon because it does not require the reference taxon to have non-zero RA for all samples. Nonetheless, it warrants further investigation to find the optimal reference taxon for the analysis which will be one of our future research topics.

**Funding**

This work was supported by NIH grants: R01GM123014, R01GM123056, P01ES022832, R01CA127334, P20GM104416, K01LM011985 and R01LM012723 and EPA grant RD-83544201.

*Conflict of Interest:* none declared.

**Appendix**

Comparison of selected genera under two randomly selected reference genera: *Akkermansia* and *Anoxybacillus*. Results for *Akkermansia* being the reference genus is also presented in Section 4.

| Table 1A. Genera associated with delivery mode and feeding type under two different reference genera. Black and green indicate positive and negative associations respectively. Genera are sorted by association strength (measured by magnitude of estimated effect size) from strongest to weakest in each category. | | | |
|---|---|---|---|
| **Genera associated with delivery mode** | | **Genera associated with feeding type** | |
| **Reference genus:** *Akkermansia* | **Reference genus:** *Anoxybacillus* | **Reference genus:** *Akkermansia* | **Reference genus:** *Anoxybacillus* |
| *Bacteroides* | *Bacteroides* | *Eubacterium* | *Eubacterium* |
| *Phascolarctobacterium* | *Parabacteroides* | *Enterococcus* | *Enterococcus* |
| *Parabacteroides* | *Phascolarctobacterium* | *Oscillospira* | *Oscillospira* |
| *Eubacterium* | *Eubacterium* | *Ruminococcus* | *Lactococcus* |
| *Megamonas* | *Collinsella* | *Lactococcus* | *Ruminococcus* |
| Collinsella | Bifidobacterium | Blautia | Blautia |
| Bifidobacterium | Sutterella | Dorea | Dorea |
| Prevotella | Prevotella | Collinsella | Collinsella |
| Ruminococcus | Limnohabitans | | Eggerthella |
| Faecalibacterium | Ruminococcus | *Haemophilus* | Parabacteroides |
| Escherichia | Megamonas | *Staphylococcus* | Granulicatella |
| Corynebacterium | Escherichia | *Serratia* | Veillonella |
| Lactobacillus | Faecalibacterium | *Propionibacterium* | Streptococcus |
| Chryseobacterium | Lactobacillus | *Citrobacter* | Lactobacillus |
| Coprobacillus | Corynebacterium | *Corynebacterium* | |
| | Acinetobacter | *Bifidobacterium* | *Staphylococcus* |
| *Clostridium* | Chryseobacterium | *Escherichia* | *Haemophilus* |
| *Veillonella* | | *Rothia* | *Serratia* |
| *Propionibacterium* | | *Peptoniphilus* | *Propionibacteriu* |
| *Serratia* | *Clostridium* | *Clostridium* | *Bifidobacterium* |
| *Atopobium* | *Veillonella* | *Acinetobacter* | *Citrobacter* |
| *Haemophilus* | *Serratia* | *Bacteroides* | *Corynebacterium* |
| *Actinomyces* | *Haemophilus* | *Pseudomonas* | *Pseudomonas* |
| *Dorea* | *Staphylococcus* | | *Rothia* |
| *Staphylococcus* | *Actinomyces* | | *Escherichia* |
| *Finegoldia* | | | *Acinetobacter* |
| *Streptococcus* | | | |